\documentstyle [sprocl,epsf,psfig]{article}

\def\be{\begin{equation}}
\def\ee{\end{equation}}
\def\bea{\begin{eqnarray}}
\def\eea{\end{eqnarray}}
\newcommand{\ket}[1]{|#1\rangle}
\newcommand{\bra}[1]{\langle #1|}
\newcommand{\Tr}{{\mbox{Tr}}}

\title{Off-diagonal mixed state phases in unitary 
evolution\footnote{To appear in "Foundations of probability 
and physics - 2", ed. by A. Khrennikov; series ``Math. Modelling 
in Physics, Engineering and Cognitive Sciences'' V\"axj\"o Univ. 
Press (2003)}}
\author{Erik Sj\"{o}qvist$^{1,}$\footnote{Electronic address: 
erik.sjoqvist@kvac.uu.se} and 
Stefan Filipp$^{2,}$\footnote{Electronic address: stefan@copilot.at}}

\address{} 

\address{$^{1}$Department of Quantum Chemistry, Uppsala University, \\  
Box 518, Se-751 20 Uppsala, Sweden}

\address{$^{2}$Institut f\"{u}r Theoretische Physik, Vienna 
University of Technology, \\ Wiedner Hauptstrasse 8-10/136, A-1040 
Vienna, Austria.}

\address{}

\begin{document}
\maketitle

\abstracts{Off-diagonal mixed state phases based upon a concept 
of orthogonality adapted to unitary evolution and a proper 
normalisation condition are introduced. Some particular 
instances are analysed and parallel transport leading to 
the off-diagonal mixed state geometric phase is delineated. 
A complete experimental realisation of the off-diagonal mixed state 
geometric phases in the qubit case using polarisation-entangled 
two-photon interferometry is proposed.}  

\section{Introduction}
When two quantal states are orthogonal their relative phase 
is indeterminate as they do not interfere. Yet, there may be a 
unitary path connecting such states and along this path there 
accumulates phase information that in part reflects the curvature 
of the subjacent state space. Thus, it seems pertinent to ask: 
is there a way to retain this particular information about the 
curvature when the path connects orthogonal states? 

This issue was settled in the pure state case only quite recently 
by Manini and Pistolesi \cite{manini00}. In essence, their idea 
may be understood by considering a unitarity $U$ and a complete 
set of orthonormalised pure states represented by the one dimensional 
projectors $\{ P_k = \ket{A_k} \bra{A_k} \}$ in terms of which one 
may define a family of off-diagonal phase factors 
\begin{eqnarray}
\gamma_{j_1j_2 \ldots j_l}^{(l)} \equiv 
\Phi \big[ \Tr \big( U P_{j_1} U P_{j_2} \ldots U P_{j_l}
\big) \big] , \ l=1,\ldots , N.  
\label{eq:man}
\end{eqnarray}
Here, $\Phi [z] = z/|z|$, $N$ is the dimensionality of the 
Hilbert space, and all $j_k$ in the set are different. These 
phase factors are manifestly independent of the choice of Hilbert 
space representatives $\{ \ket{A_k} \}$, and hence measurable in 
principle. Furthermore, they are independent of cyclic permutations 
of the indexes $j_{1},j_{2}, \ldots j_{l}$, they contain the 
standard Pancharatnam phase factor \cite{pancharatnam56} as  
$l=1$, they reflect the curvature of the subjacent state space 
if $U$ parallel transports $\{ \ket{A_k} \}$, and experimental 
test of the $l=2$ case has been reported for spin polarised 
neutrons \cite{hasegawa01}. 

In this paper, we wish to elaborate on the concept of off-diagonal 
phases for mixed states proposed in \cite{filipp02}. For parallel 
transporting unitarities this defines a family of off-diagonal 
mixed state geometric phases that reflects the geometry of the 
subjacent state space. These phases extend the concept of mixed state 
phase in \cite{sjoqvist00} to cases where the latter is undefined.  

Unitary maps of a complete orthonormal set of states are characterised 
and the concomitant concept of quantum parallel transport is
delineated in the next section. We propose an operationally natural
notion of orthogonality adapted to unitarily connected density
matrices in section {\bf 3}. This is used for the off-diagonal mixed
state geometric phase proposed in section {\bf 4}. Section {\bf 5} 
contains some explicit examples and a complete experimental realisation 
of the off-diagonal mixed state geometric phases in the qubit case is 
proposed in section {\bf 6}. The paper ends with the conclusions.

\section{$U(N)$, $SU(N)$, and quantum parallel transport}
Any complete orthonormal basis of a finite dimensional Hilbert 
space is mapped unitarily to another complete orthonormal basis. 
Here, we provide some general remarks on unitary maps of such 
bases that are pertinent in the context of off-diagonal 
mixed state phases. 

Consider Hilbert space ${\cal H}$ of finite dimension $N$. 
Any unitary map acting on ${\cal H}$ can be decomposed as 
$U(N) = U(1) \times SU(N)$. As is clear from Eq. (\ref{eq:man}), 
the $U(1)$ part factors out and contributes a factor 
$\big[U(1)\big]^l$ to $\gamma^{(l)}$. 

We may also consider parallel transporting unitarities. Indeed, 
parallel transport of a pure quantum state, probably first put 
forward in \cite{simon83}, plays an important role in the theory 
of geometric phases as in this case the dynamical contributions 
along the path are assured to vanish. In the context of off-diagonal 
phases, it proves useful to extend this and consider parallel 
transport of a set of orthonormal pure states.

Consider a complete orthonormal basis $\ket{A_k}$ of ${\cal H}$. 
A continuous one-parameter family of unitarities $U(s)$ is said 
to parallel transport the basis $\ket{A_k}$ if it fulfils 
\begin{equation}
\bra{A_k} U^{\dagger} \dot{U} \ket{A_k} = 0, \ \forall k , 
\label{eq:pc}
\end{equation}
which is equivalent to having no local accumulation of phase 
along the unitary path for each $\ket{A_k}$. We may further 
notice that any unitarity $U(s)$ may be written as  
\begin{equation}
U(s) = {\cal P} \exp \left( -i \int_{0}^{s} J(s') ds' \right) , 
\end{equation}
$J(s)$ being Hermitian and ${\cal P}$ is path ordering. 
Thus, equivalent to Eq. (\ref{eq:pc}) is 
\begin{equation}
\bra{A_k} J(s) \ket{A_k} = 0, \ \forall k, 
\label{eq:pcj}
\end{equation}
which entails that $J(s)$ has to be off-diagonal in the parallel 
transported basis and therefore traceless in any basis. Thus, 
$U \in SU(N)$ is a necessary (but not sufficient) condition 
for $U$ being parallel transporting a complete basis\footnote{Less 
restrictive conditions may be put on $U$ by considering parallel 
transport in a fixed $K<N$ dimensional subspace of ${\cal H}$. 
In such a case $U(N) = SU(K) \times U(N-K)$, where $SU(K)$ 
parallel transport some basis of this subspace. Such unitarites 
are useful when considering rank $<N$ density operators.}.

The conditions in (\ref{eq:pc}) or equivalently in (\ref{eq:pcj}) 
define a nontrivial fibre bundle with structure group being 
isomorphic to the $N$ torus. In the context of mixed states, 
this is the relevant bundle structure in \cite{sjoqvist00}. 
Uhlmann \cite{uhlmann86} has provided another concept of mixed 
state parallel transport that defines a $U(N)$ bundle and differs 
both conceptually and physically \cite{ericsson02} from that of 
\cite{sjoqvist00}. In this report, we focus on the extension of 
\cite{sjoqvist00} to the off-diagonal case.

\section{Orthogonality} 
Generalisation of the off-diagonal phases to the mixed state
case requires an appropriate notion of ``orthogonality'' between
unitarily connected density matrices. In this section, we propose 
a simple definition of this based upon interference.

In order to develop this idea, let us first suppose $\ket{A}$ and 
$\ket{B}$ are Hilbert space representatives of two arbitrary 
pure quantal states $A$ and $B$, and assume further that 
$\ket{A}$ is exposed to the variable $U(1)$ shift $e^{i\chi}$. 
The resulting interference pattern obtained in superposition 
is determined by the intensity profile 
\begin{equation}
{\cal I} = \Big| e^{i\chi}|A\rangle + |B\rangle \Big|^{2} = 
2 + 2 |\langle A|B \rangle| 
\cos [ \chi - \arg \langle A|B \rangle ] . 
\label{eq:pureinterfer}
\end{equation}  
The key point here is to note that $A$ and $B$ are orthogonal 
if and only if ${\cal I}$ is independent of $\chi$. 

To extend this idea to the mixed state case, consider a pair 
of unitarily connected density operators    
\begin{equation}
\rho_{A} = \sum_{k} \lambda_{k} |A_{k} \rangle \langle A_{k}| 
\longrightarrow \rho_{B} = 
\sum_{k} \lambda_{k} |B_{k} \rangle \langle B_{k}| , 
\end{equation}
where each $|B_{k}\rangle = U|A_{k} \rangle$. Evidently, 
each such orthonormal pure state component of the density 
operator contributes to the interference according to Eq. 
(\ref{eq:pureinterfer}). Thus, the total intensity 
profile becomes 
\begin{eqnarray} 
{\cal I} & = & \sum_{k} \lambda_{k} \Big| e^{i\chi}|A_{k}\rangle + 
|B_{k}\rangle \Big|^{2} 
\nonumber \\ 
 & = & 2 + 2 \sum_{k} \lambda_{k} |\langle A_{k}|B_{k} \rangle| 
\cos [ \chi - \arg \langle A_{k}|B_{k} \rangle ] , 
\end{eqnarray}  
where we have used that the $\lambda$'s sum up to unity.
Following the above pure state case, we say that $\rho_A$ and 
$\rho_B$ are orthogonal if and only if ${\cal I}$ is independent 
of $\chi$ for all Hilbert space representatives $\{ \ket{A_k} \}$ 
and $\{ \ket{B_k} \}$ of the eigenstates of $\rho_A$ and $\rho_B$, 
respectively. It follows that $\rho_A \perp \rho_B$ if and only 
if $\langle A_k\ket{B_k} = 0, \ \forall k$.

For an $N$ dimensional Hilbert space ${\cal H}$, we may 
generate a set of $N$ mutually orthogonal density operators 
as follows.  Assume there is a unitary operator $U_g$ such that 
$\ket{A_n} = \big( U_g \big)^{n-1} \ket{A_1}, \ n=1,\ldots ,N,$ 
is a complete orthonormal basis of ${\cal H}$. Explicitly, we 
may write 
\begin{equation} 
U_g = \ket{A_1} \bra{A_N} + \ket{A_N} \bra{A_{N-1}} + 
\ldots \ket{A_2} \bra{A_1} . 
\end{equation} 
Now, if $\rho_{1} \ket{A_k} = \lambda_k \ket{A_k}$ then 
\begin{equation} 
\rho_{n} = \big( U_g \big)^{n-1} \rho_{1} \big( U_g^{\dagger} 
\big)^{n-1} , \ n=1,\ldots , N
\end{equation}
is a set of mutually orthogonal density operators. Explicitly, 
this entails that 
\begin{eqnarray} 
\rho_{1} & = & \lambda_{1} \ket{A_1} \bra{A_1} + 
\lambda_{2} \ket{A_2} \bra{A_2} + \ldots + 
\lambda_{N} \ket{A_N} \bra{A_N} , 
\nonumber \\ 
\rho_{2} & = & \lambda_{1} \ket{A_2} \bra{A_2} + 
\lambda_{2} \ket{A_3} \bra{A_3} + \ldots +  
\lambda_{N} \ket{A_1} \bra{A_1} , 
\nonumber \\ 
 & \ldots , & 
\nonumber \\ 
\rho_{N} & = & \lambda_{1} \ket{A_N} \bra{A_N} + 
\lambda_{2} \ket{A_1} \bra{A_1} + \ldots + 
\lambda_{N} \ket{A_{N-1}} \bra{A_{N-1}} . 
\label{eq:Nstates}
\end{eqnarray}

\section{Off-diagonal mixed states phases}
In this section, we propose the off-diagonal phases for mixed 
states, based upon the concept of orthogonality described above. 
To do this, we only need to determine how the mutually orthogonal 
density operators should appear in the trace. This may be resolved 
by noting that:  
\begin{itemize}
\item For $U=U_g^{\dagger}$ that permutes  
$\ket{A_N}\rightarrow\ket{A_{N-1}}\rightarrow \ldots \rightarrow 
\ket{A_1} \rightarrow\ket{A_N}$, we have 
\begin{equation}
\Tr \big( U_g^{\dagger} P_1 
U_g^{\dagger} P_2 \ldots U_g^{\dagger} P_N \big) = 1,
\end{equation}
due to normalisation $\bra{A_k} A_k \rangle =1$, $\forall k$. 
\item The ansatz $\rho_k^{p/q}$ reduces to $P_k$ in the limit 
of pure states, if $p$ and $q$ are integers. Note here that 
$\rho_k^{p/q}$ is well-defined as $\rho_k \geq 0$. 
\end{itemize}
Replace $P_k$ with $\rho_k^{p/q}$ in Eq. (\ref{eq:man}) and consider 
$l=N$ with $j_k = k$, $\forall k$. For $U=U_g^{\dagger}$ we have 
\begin{eqnarray} 
\Tr \big( U_g^{\dagger} \rho_1^{p/q} U_g^{\dagger} \rho_2^{p/q} \ldots 
U_g^{\dagger} \rho_N^{p/q} \big) = \Tr \big( \rho_1^{Np/q} \big) , 
\end{eqnarray} 
where we have used that $\big( U_g \rho U_g^{\dagger} 
\big)^{p/q} = U_g \rho^{p/q} U_g^{\dagger}$ for any integers $p$ and 
$q$. Normalisation yields $p=1$ and $q=N$. Now, for general 
$l\leq N$, assume that $p$ and $q$ are completely determined by $l$, 
we may write $p(l) = \sum_m a_m l^m$ and $q(l) = \sum_m b_m l^m$, 
with $l-$independent integer coefficients $\{ a_m \}$ and $\{ b_m \}$. 
From $p(N)=1$ and $q(N)=N$ for any $N$, we obtain $a_m = \delta_{m,0}$ 
and $b_m = \delta_{m,1}$. Thus, $p=1$ and $q=l$. 

We are now ready to state our main result: the off-diagonal mixed 
state phase for an ordered set of $l\leq N$ mutually orthogonal 
density matrices $\rho_{j_k}$, $k=1, \ldots ,l$, transported by 
$U$ is naturally given by 
\begin{equation}
\gamma_{\rho_{j_1}\rho_{j_2}\ldots\rho_{j_l}}^{(l)} 
\equiv \Phi \big[ \Tr \big( U \sqrt[l]{\rho_{j_1}} 
U \sqrt[l]{\rho_{j_2}} \ldots U \sqrt[l]{\rho_{j_l}} \big) \big] .   
\label{eq:genoffdiag}
\end{equation}
This is manifestly gauge invariant and independent of cyclic 
permutations of the indexes $j_{1},j_{2}, \ldots j_{l}$. By 
construction it reduces to Eq. (\ref{eq:man}) in the limit 
of pure states. If $U$ is parallel transporting then Eq. 
(\ref{eq:genoffdiag}) defines a family of off-diagonal  
mixed state geometric phases. Furthermore, just as in the 
pure state case, any $U(1)$ component of $U$ contributes 
here with a factor $\big[ U(1) \big]^{l}$ to $\gamma^{(l)}$. 

The mixed state phase 
\begin{equation} 
\gamma_{\rho_{j_1}}^{(1)} = 
\Phi \big[ \Tr \big( U \rho_{j_1} \big) \big]  
\label{eq:mixedl1}
\end{equation}
proposed in \cite{sjoqvist00} may be seen as a natural 
consequence of this general framework if we put $l=1$. 
In section {\bf 6} we propose an experimental realisation 
of the $l=2$ case 
\begin{equation}
\gamma_{\rho_{j_1} \rho_{j_2}}^{(2)} =  
\Phi \big[ \Tr \big( U \sqrt{\rho_{j_1}} U 
\sqrt{\rho_{j_2}} \big) \big]  
\label{eq:mixedl2}
\end{equation}
in polarisation-entangled two-photon interferometry. 

Note that if $\rho_1$ and $U_g$ commutes for some $n$ then 
$\rho_n = \rho_1$. This may happen for $n\neq 1$ if and only 
if $\rho_1 = I/N$, $I$ being the identity operator on 
${\cal H}$, i.e. for the maximally mixed state. In such a 
case $\rho_1 = \rho_2 = \ldots = \rho_N$ and all $\gamma^{(l)}$ 
solely reflect properties of $U$. 

\section{Examples}
In the qubit case, consider an $SU(2)$ operator  
\begin{equation} 
U = U_{11} \ket{A_1} \bra{A_1} + U_{12} \ket{A_1} \bra{A_2} + 
U_{21} \ket{A_2} \bra{A_1} + U_{22} \ket{A_2} \bra{A_2} 
\end{equation}
in the common eigenbasis of the mutually orthogonal $\rho_1$ and 
$\rho_2$. This yields 
\begin{eqnarray} 
\Tr \big( U \rho_{1} \big) & = & 
\eta \big( \lambda_1 e^{i\alpha} + \lambda_2 e^{-i\alpha} \big)  
\nonumber \\ 
\Tr \big( U \rho_{2} \big) & = & 
\eta \big( \lambda_1 e^{-i\alpha} + \lambda_2 e^{i\alpha} \big)  
\nonumber \\ 
\Tr \big( U \sqrt{\rho_1} U 
\sqrt{\rho_2} \big) & = & 2 \eta^2 \sqrt{\lambda_1\lambda_2} 
\cos 2 \alpha -1 + \eta^2 ,  
\end{eqnarray}
where we have used $U_{11} = U_{22}^{\ast} = \eta e^{i\alpha}$ 
and $U_{12} U_{21} = - \det U + U_{11} U_{22} = -1+\eta^2$ for 
$SU(2)$. If $U$ is parallel transporting then $\alpha$ is the 
geodesically closed solid angle enclosed by the Bloch vector. 

In the nondegenerate case $\lambda_1 \neq \lambda_2$, the $l=1$ 
phases are indeterminate only for $\eta =0$, for which the 
$l=2$ phase is well-defined since $\Tr \big( U \sqrt{\rho_1} 
U \sqrt{\rho_2} \big) = -1$. In the degenerate case $\lambda_1 = 
\lambda_2 = \frac{1}{2}$, $\Tr \big( U \rho_1 \big)$ and 
$\Tr \big( U \rho_2 \big)$ have additional nodal points as 
discussed in \cite{bhandari02,anandan02}. These occur whenever 
$\cos \alpha = 0$, at which angles $\cos 2 \alpha = -1$ and we 
again have $\Tr \big( U \sqrt{\rho_1} U \sqrt{\rho_2} \big) = -1$.
Thus, $\gamma^{(1)}$ and $\gamma^{(2)}$ never become indeterminate 
simultaneously and thus provide a complete phase characterisation 
of the qubit case. 

The complexity of the analysis increases rapidly with $N$. For 
simplicity we therefore focus on two important special kinds of 
unitarities for $N\geq 2$: 
\begin{itemize}
\item[(i)] Diagonal unitarities  
\begin{equation}
U_d = \sum_{k=1}^{N} U_{kk} \ket{A_k} \bra{A_k} .
\end{equation}
Here, all $|U_{kk}| =1$ and for $SU(N)$ we have $U_{11} U_{22} 
\ldots U_{NN} = +1$. If $U_d$ is parallel transporting then 
$U_{kk} = \gamma_k^{(1)}$, $\gamma_k^{(1)}$ being the cyclic geometric 
phase factor of the pure state $A_k$. 
\item[(ii)] Permuting unitarites, which may be written as 
\begin{equation}
U_p = U_{12} \ket{A_1} \bra{A_2} + U_{23} \ket{A_2} \bra{A_3} + 
\ldots + U_{N1} \ket{A_N} \bra{A_1} . 
\end{equation}
Here, all $|U_{12}| = |U_{23}| = \ldots = |U_{N1}| =1$ and 
in the case of $SU(N)$ we have $U_{12} U_{23} \ldots U_{N1} = 
(-1)^{N-1}$. 
\end{itemize}
These two cases can be considered as extremes in the sense 
that (i) corresponds to cyclic evolution of the common 
eigenstates of the $\rho$'s while (ii) is a particular 
instance where each of these eigenstates evolves into an 
orthogonal state. Combinations of these two extremes 
are discussed in \cite{filipp02}. 

First, let us consider the diagonal case. We have 
\begin{eqnarray} 
\Tr \big( U_d \sqrt[l]{\rho_{j_1}} \ldots 
U_d \sqrt[l]{\rho_{j_l}} \big) = 
\sum_{k=1}^N \big( U_{kk} \big)^l 
\sqrt[l]{\lambda_{k_1} \ldots \lambda_{k_l}} . 
\label{eq:diagonal}
\end{eqnarray}
As each term contains precisely $l$ $\lambda$'s, it follows 
that all $\Tr \big( U_d \sqrt[l]{\rho_{j_1}} \ldots U_d 
\sqrt[l]{\rho_{j_l}} \big)$ must vanish if $l>$ rank of the $\rho$'s.  

In the permutation case we first notice that 
\begin{eqnarray} 
U_p \sqrt[l]{\rho_k} & = & 
\sqrt[l]{\lambda_1} U_{k-1,k} \ket{A_{k-1}} \bra{A_{k}} + \ldots + 
\sqrt[l]{\lambda_{N-k+1}} U_{N-1,N} \ket{A_{N-1}} \bra{A_{N}} 
\nonumber \\ 
 & & + \sqrt[l]{\lambda_{N-k+2}} U_{N1} \ket{A_{N}} \bra{A_{1}} + 
\ldots + \sqrt[l]{\lambda_N} U_{k-2,k-1} \ket{A_{k-2}} \bra{A_{k-1}}  
\nonumber \\  
\label{eq:permult}
\end{eqnarray}
with the identifications $U_{01} \equiv U_{N1}$ and $\ket{A_0} 
\equiv \ket{A_N}$. Thus, multiplying $l$ such factors results in 
a sum of operators of the form $\ket{A_N} \bra{A_{N-l}}, \ldots ,
\ket{A_1} \bra{A_{N-l+1}}$, whose trace may be nonvanishing  
only if $l=N$. Thus, all $\gamma^{l<N}$ are indeterminate. Furthermore, 
upon multiplication of $N$ factors $U_p \sqrt[N]{\rho_k}$, it 
follows from Eq. (\ref{eq:permult}) that $\Tr \big( U_p 
\sqrt[N]{\rho_{j_1}} \ldots U_p \sqrt[N]{\rho_{j_l}} \big)$ 
results in a sum where each term contains the product 
$U_{12}U_{23} \ldots U_{N1}$. For $SU(N)$, this implies 
that we may write  
\begin{equation} 
\Tr \big( U_p \sqrt[N]{\rho_{j_1}} \ldots 
U_p \sqrt[N]{\rho_{j_l}} \big) = (-1)^{N-1}
f_{\rho_{j_1}\ldots\rho_{j_N}}^{(N)} (\lambda_1,\ldots,\lambda_N) , 
\label{eq:permutation}
\end{equation}
where each $f^{(N)}$ is determined by the sequence of $\rho$'s. 

The $f$'s in Eq. (\ref{eq:permutation}) have some interesting 
properties. First, it can be seen that 
\begin{equation} 
f_{\rho_{1}\ldots\rho_{N}}^{(N)} = 1, \ \forall N. 
\end{equation}
Thus, there exist at least one well-defined off-diagonal mixed 
state phase for $U_p$, independent of the rank of the $\rho$'s. 
Secondly, we have 
\begin{equation} 
f_{\rho_{j_1}\ldots\rho_{j_N}}^{(N)} \geq 0, \ \forall 
j_1, \ldots ,j_N . 
\end{equation}
This implies that the off-diagonal mixed state phases for $U_p$ 
are completely determined by the dimension of the Hilbert space 
${\cal H}$. Indeed, for sequences where $f^{(N)} \neq 0$ we have 
\begin{eqnarray} 
\gamma^{(N)} & = & -1, \ {\textrm{if dim(${\cal H}$) even,}} 
\nonumber \\ 
\gamma^{(N)} & = & +1, \ {\textrm{if dim(${\cal H}$) odd.}}
\end{eqnarray} 

\section{Two-photon experiment}
When considering the issue of experimental realisation of the
off-diagonal mixed state phases we immediately encounter a problem: 
how do we experimentally implement the $l$th root of density operators?
Fortunately, this may be resolved in the $l=2$ case in the sense of
purification, i.e. by adding an ancilla system in a certain way. 
Here, we demonstrate this in the qubit case in terms of an explicit 
experiment for polarisation-entangled photon pairs. The set up is 
sketched in Fig. 1. 

Consider an ensemble of linearly polarised photons with 
polarisation degree $r$. In the horizontal-vertical $(h-v)$ 
basis, there are two possible unitarily equivalent and orthogonal 
representions of the ensemble in terms of the density operators 
\begin{eqnarray} 
\rho_1 & = & \frac{1+r}{2} \ket{h} \bra{h} + 
\frac{1-r}{2} \ket{v} \bra{v} ,  
\nonumber \\ 
\rho_2 & = & \frac{1-r}{2} \ket{h} \bra{h} + 
\frac{1+r}{2} \ket{v} \bra{v} . 
\end{eqnarray}
A purification of any of these density operators may be achieved by 
adding an ancilla photon in such a way that the photon pair is in 
a pure polarisation state whose partial trace over the ancilla is 
the density operator. The polarisation-entangled state 
\begin{eqnarray}
\ket{\Psi_1} & = & \sqrt{\frac{1}{2}(1+r)} \ket{h} 
\otimes \ket{h} + \sqrt{\frac{1}{2}(1-r)} \ket{v} 
\otimes \ket{v} , 
\end{eqnarray} 
which has been demonstrated in \cite{kwiat99}, is an example of a 
purification of $\rho_1$. 

For simplicity, we consider unitarities that rotate linear
polarisation states along great circles an angle $\beta$ on 
the Poincar\'{e} sphere. This amounts to
\begin{equation}
U (\beta,\theta) = 
\exp \Big( -i\beta \Big[ \cos \theta \big( \ket{h}\bra{v} + 
\ket{v}\bra{h} \big) + 
\sin \theta \big( -i\ket{h}\bra{v} + i\ket{v} \bra{h} \big) 
\Big] \Big) ,
\label{eq:polarrotation}   
\end{equation}
which fulfils the parallel transport conditions in Eqs. (\ref{eq:pc}) 
and (\ref{eq:pcj}) with respect to the $h-v$ basis. $U(\beta,\pi/2)$ 
takes linear polarisation into linear polarisation with plane of 
polarisation rotated an angle $\beta$. An important special case 
is the polarisation flip $F=U(\pi /2,\pi /2)$ that connects $\rho_1$ 
and $\rho_2$. Furthermore, for $\theta =0$ and $\beta = \pi /4$, 
circular polarisation states are obtained. 

We shall now demonstrate how purification may be used in the set up
shown in Fig. 1 to test $\gamma^{(1)}$ and $\gamma^{(2)}$ for
$\rho_1$ and $\rho_2$ in the case of $U=U(\beta,\theta)$. With 
$\ket{\Psi_1}$ as input, the intensity detected in coincidence is
\cite{hessmo00}
\begin{eqnarray}
{\cal I} & = & \big| U_s \otimes U_a \ket{\Psi_1} +  
V_s \otimes V_a \ket{\Psi_1} \big|^{2} 
\nonumber \\ 
 & = &  2 + 2 \Re \Big[ \bra{\Psi_1} U_s^{\dagger} V_s \otimes 
U_a^{\dagger} V_a \ket{\Psi_1} \Big] ,  
\end{eqnarray} 
where we have used that simultaneous detection occurs only the 
photons both either took the shorter path or the longer path 
(assuming sufficiently short coincidence window). By appropriate 
choices of the unitarities shown in Fig. 1, we may obtain the 
$\gamma$'s as follows. 
\begin{itemize}
\item[$\bullet$ $\gamma_{\rho_1}^{(1)}$:] Choose $U_s = e^{i\chi}$, 
$V_s = U(\beta,\theta)$, and $U_a = V_a = I$ yielding 
\begin{eqnarray} 
\bra{\Psi_1} U_s^{\dagger} V_s \otimes 
U_a^{\dagger} V_a \ket{\Psi_1} & = & 
e^{-i\chi} \Tr \big[  U(\beta,\theta) \otimes I 
\ket{\Psi_1} \bra{\Psi_1}\big]  
\nonumber \\ 
 & = & e^{-i\chi} \Tr_s \big[ U(\beta,\theta) \rho_1 \big] ,  
\end{eqnarray} 
where we have used that $\Tr_a \big[ \ket{\Psi_1} \bra{\Psi_1} \big] = 
\rho_1$. Thus, $\arg \gamma_{\rho_1}^{(1)}$ is the shift obtained 
by variation of $\chi$. Explicit calculation for $U(\beta,\theta)$ 
in Eq. (\ref{eq:polarrotation}) entails that $\gamma_{\rho_1}^{(1)}$
is real-valued and changes sign at $\beta = (j+\frac{1}{2})\pi$, $j$ 
integer, corresponding to a sequence of phase jumps of $\pi$. 
\item[$\bullet$ $\gamma_{\rho_2}^{(1)}$:] Choose $U_s = e^{i\chi}F$, 
$V_s = U(\beta,\theta)F$, and $U_a = V_a = I$ yielding 
\begin{eqnarray} 
\bra{\Psi_1} U_s^{\dagger} V_s \otimes 
U_a^{\dagger} V_a \ket{\Psi_1} & = & 
e^{-i\chi} \Tr \big[ U(\beta,\theta) F \otimes I 
\ket{\Psi_1} \bra{\Psi_1} F^{\dagger} \big]  
\nonumber \\ 
 & = & e^{-i\chi} \Tr_s \big[ U(\beta,\theta) \rho_2 \big] , 
\end{eqnarray} 
where we have used that $\Tr_a \big[ F \ket{\Psi_1} \bra{\Psi_1} 
F^{\dagger} \big] = \rho_2$. Thus, $\arg \gamma_{\rho_2}^{(1)}$ is the 
shift obtained by variation of $\chi$. Also $\gamma_{\rho_2}^{(1)}$
is real-valued and changes sign at $\beta = (j+\frac{1}{2})\pi$, 
$j$ integer, for $U(\beta,\theta)$. 
\item[$\bullet$ $\gamma_{\rho_1 \rho_2}^{(2)}$:] Choose $U_s = e^{i\chi} F$, 
$V_s = U(\beta,\theta)$, $U_a = F$, and $V_a = U(\beta,-\theta)$ 
yielding 
\begin{eqnarray} 
\bra{\Psi_1} U_s^{\dagger} V_s \otimes 
U_a^{\dagger} V_a \ket{\Psi_1} & = & 
e^{-i\chi} \Tr \big[ U(\beta,\theta) \otimes U(\beta,-\theta) 
\nonumber \\ 
 & & \times \ket{\Psi_1} \bra{\Psi_1} 
F^{\dagger} \otimes F^{\dagger} \big] . 
\end{eqnarray} 
Explicit calculation yields 
\begin{eqnarray} 
\Tr \big[ U(\beta,\theta) \otimes U(\beta,-\theta) \ket{\Psi_1} 
\bra{\Psi_1} F^{\dagger} \otimes F^{\dagger} \big] & = &  
\Tr_s \big[ U \sqrt{\rho_1} U \sqrt{\rho_2} \big] .  
\end{eqnarray}
Thus, $\arg \gamma_{\rho_1\rho_2}^{(2)}$ is the shift obtained 
by variation of $\chi$. Furthermore, we may compute the expected 
output as  
\begin{equation} 
\Tr_s \big[ \sqrt{\rho_1} U \sqrt{\rho_2} U \big] = 
\sqrt{1-r^2} \cos^{2} \beta - \sin^{2} \beta , 
\end{equation}
which is independent of $\theta$ and can be positive and negative 
for $r\neq 1$ depending upon $\beta$. Such an experiment would 
test that the off-diagonal geometric phase is either $0$ or $\pi$ 
for mixed qubit states. 
\end{itemize}

\section{Conclusions}
The concept of geometric phase has recently been extended to cases
where the standard definition breaks down. Such cases occur if a
unitarity connects orthogonal pure states \cite{manini00} or if a
unitarity connects mixed states \cite{sjoqvist00}. Here we have
reported on a unification of these extensions: the off-diagonal mixed
state phase that also covers situations where mixed states do not
interfere in the sense of \cite{sjoqvist00}. Although the present
off-diagonal mixed state phases are properties of the system (they are
expressed solely in terms of a set of density operators pertaining to
the system) experimental realisations thereof seem to require control 
and measurement of one or possibly several additional ancilla systems. 
We have proposed an explicit Franson interferometer set up for
polarisation-entangled photon pairs as a complete experimental
realisation of the off-diagonal mixed state phase in the qubit
case. Such an experiment would in particular demonstrate a 
nontrivial sign change property of the off-diagonal qubit phase 
that is associated with the mixed state case. We hope that the 
ideas reported here would trigger new experimental tests as well 
as to further theoretical considerations of off-diagonal phases. 

\section*{Acknowledgments} 
The work by E.S. was supported by the Swedish Research Council.

\newpage 
\section*{Figure Captions} 
Fig.1. Franson set up for polarisation-entangled photon pairs. 
In the longer arms, the system and ancilla photons are exposed to 
the unitarities $U_s$ and $U_a$, respectively, and similarly 
$V_s$ and $V_a$ in the shorter arms. 
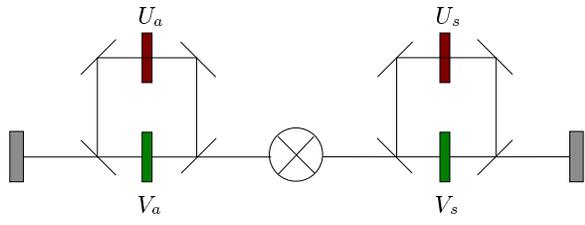
\end{document}